\documentclass[prl,twocolumn,showpacs,preprintnumbers,amsmath,amssymb]{revtex4}

\usepackage{graphicx}% Include figure files
\usepackage{dcolumn}% Align table columns on decimal point
\usepackage{bm}% bold math

\begin{document}

\preprint{APS/123-QED}

\title{Correlation between $T_c$ and anisotropic scattering in Tl$_2$Ba$_2$CuO$_{6+\delta}$}

\author{M. Abdel-Jawad}
\author{J. G. Analytis}

\affiliation{H. H. Wills Physics Laboratory, University of Bristol, Bristol BS8 1TL, United Kingdom}

\author{L. Balicas}

\affiliation{National High Magnetic Field Laboratory, Florida State University, Tallahassee, Florida 32306, U.S.A.}

\author{A. Carrington}
\affiliation{H. H. Wills Physics Laboratory, University of Bristol, Bristol BS8 1TL, United Kingdom}

\author{J. P. H. Charmant}
\affiliation{The School of Chemistry, University of Bristol, Bristol BS8 1TS, United Kingdom }

\author{M. M. J. French}
\affiliation{H. H. Wills Physics Laboratory, University of Bristol, Bristol BS8 1TL, United Kingdom}

\author{N. E. Hussey}
\affiliation{H. H. Wills Physics Laboratory, University of Bristol, Bristol BS8 1TL, United Kingdom}

\date{\today}

\begin{abstract}
Angle-dependent magnetoresistance measurements are used to determine the isotropic and anisotropic components of the
transport scattering rate in overdoped Tl$_2$Ba$_2$CuO$_{6+\delta}$ for a range of $T_c$ values between 15K and 35K.
The size of the anisotropic scattering term is found to scale linearly with $T_c$, establishing a link between the
superconducting and normal state physics. Comparison with results from angle resolved photoemission spectroscopy
indicates that the transport and quasiparticle lifetimes are distinct.
\end{abstract}

\pacs{74.72.Jt, 75.47.-m}

\maketitle

Understanding the normal state is regarded as a key step in resolving the problem of high temperature superconductivity
(HTSC) yet establishing any clear correlation between the two has proved difficult. Most empirical correlations to
date, such as the Uemura plot \cite{Uemura} and the linear scaling of the magnetic resonance mode energy with $T_c$
{\cite{ResonanceMode}, are associated with the superconductivity. Homes' law, linking $T_c$ with the product of the
superfluid density and the dc conductivity (at $T_c$) \cite{Homes}, is a rare example of a correlation linking the two
states, though this also could be viewed as a consequence of superconducting (SC) gap formation rather than a normal
state property \cite{HusseyReview}. Transport properties, particularly for current in-plane, appear very much tied to
the $T_c $ parabola (for a review see \cite{HusseyReview}) but any direct correlation between transport and
superconductivity has not yet been found. Finally, the precise doping dependence of the pseudogap and its relation to
$T_c$ remains a controversy \cite{TallonLoram}.

Here we identify a new correlation between the normal and superconducting states using a bulk transport probe, namely
interlayer angle-dependent magnetoresistance (ADMR). \cite{Yamaji} ADMR have provided detailed Fermi surface (FS)
information for a variety of one- and two-dimensional (2D) metals \cite{KartsovnikReview, BergemannAP, Hussey03}.
Recently the technique was extended to incorporate basal-plane anisotropy and to reveal the temperature $T$- and
momentum ({\bf k}-) dependence of the scattering rate $\Gamma$($T$,{\bf k}) in heavily overdoped (OD)
Tl$_2$Ba$_2$CuO$_{6+\delta}$ (Tl2201) \cite{Majed06}. There $\Gamma$($T$,{\bf k}) was found to consist of two
components, one isotropic and quadratic in $T$, the other anisotropic, maximal near the saddle points at ($\pi$, 0) and
proportional to $T$.

In this Letter, ADMR measurements at $T$ = 40K and magnetic field $\mu_0H$ = 45Tesla are compared for a number of OD
Tl2201 crystals with $T_c$ values between 15K and 35K. The strength of the anisotropic scattering extracted from the
analysis is found to scale linearly with $T_c$, {\it appearing to extrapolate to zero at the doping level where
superconductivity vanishes}. This finding implies that the anisotropic scattering mechanism is intimately related to
the mechanism of HTSC. In marked contrast to recent results from angle resolved photoemission spectroscopy (ARPES)
\cite{Zhou04, Plate05}, no sign reversal of the anisotropy in the quasiparticle lifetime is inferred. Finally our
results shed new light on the doping evolution of both $\rho_{ab}$($T$) and $R_{\rm H}$($T$) in OD cuprates.

%\section{Experimental method and simulation procedure}

For this study a total of six tetragonal self-flux grown crystals (typical dimensions 0.3 $\times$ 0.3 $\times$
0.03mm$^3$) were annealed at temperatures 300$^{\rm o}$C $\leq T \leq$ 600$^{\rm o}$C in flowing O$_2$ and mounted in a
$c$-axis quasi-Montgomery configuration. The ADMR were measured on a two-axis rotator in the 45T hybrid magnet at the
NHMFL in Florida using a conventional four-probe ac lock-in technique. The orientation of the crystal faces was indexed
for a number of crystals using a single crystal x-ray diffractometer.

\begin{figure*}
\includegraphics[width=6.0cm,angle=270,keepaspectratio=true]{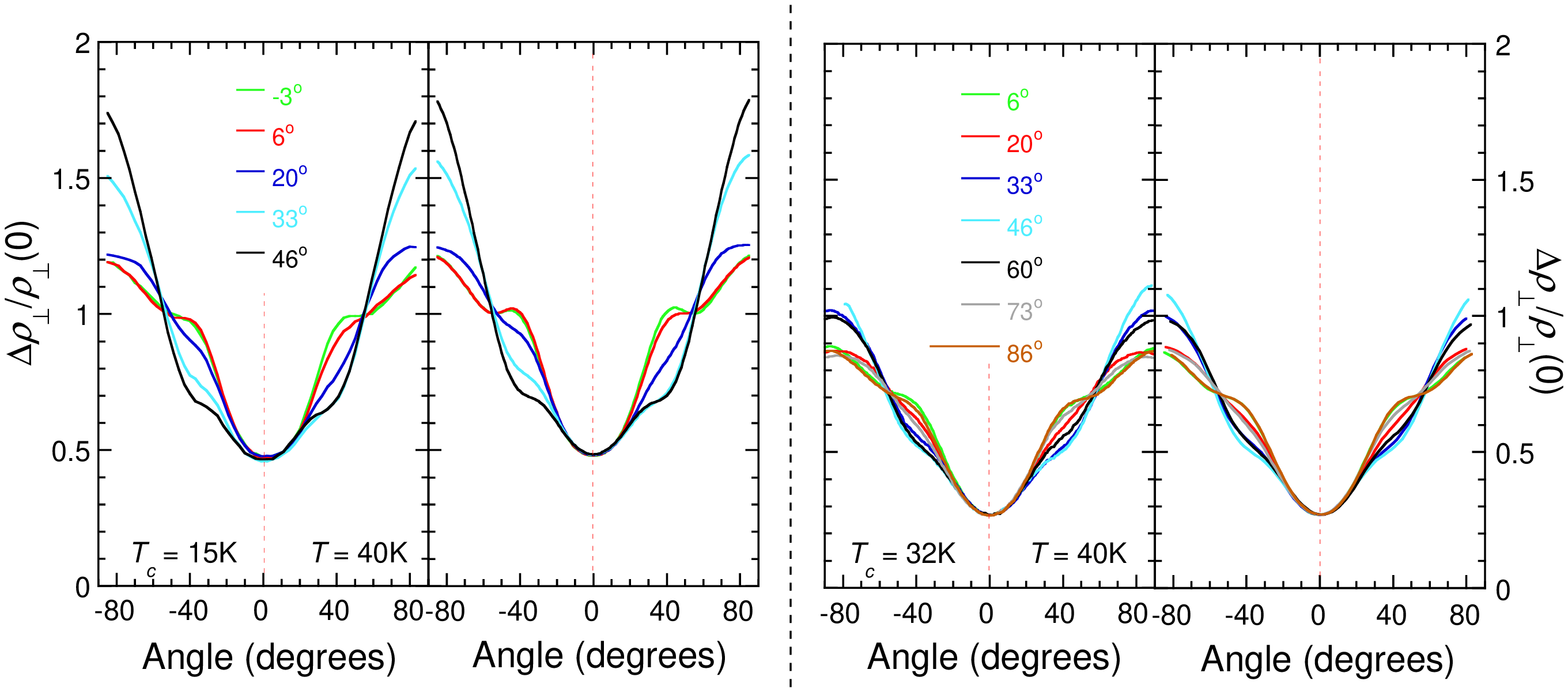}
\caption{(Color online) Angle-dependent magnetoresistance sweeps for two overdoped Tl2201 single crystals with $T_c$ of
15K (Tl15Kb - left side) and 32K (Tl32K - right side) measured at 40K and 45T. The least-square fits to the data are
displayed on the right of each plot. The labels refer to the azimuthal angle of the magnetic field with respect to the
Cu-O-Cu bond direction. \label{fig1}}
\end{figure*}

The first and third panels of Fig. \ref{fig1} show polar ADMR data $\Delta\rho_{\bot}/\rho_{\bot}$(0) (normalized to
their zero-field value) at different azimuthal angles $\phi$ (relative to the Cu-O-Cu bond direction) for two crystals
Tl15Kb and Tl32K (the numbers relate to their $T_c$ values). Though the two sets of data look similar, there are some
key differences \cite{asymmetry}. Firstly $\Delta\rho_{\bot}/\rho_{\bot}$(0) is significantly larger for the lower
$T_c$ crystal. Secondly, for polar angles $|\theta| \sim \pi/2$, Tl32K shows much less $\phi$-dependence and finally,
near {\bf H}$\|c$ ($\theta$ = 0), the Tl15Kb curves are more rounded. As shown below, these features are caused by the
higher $T_c$, less overdoped sample(s) possessing a significantly larger basal-plane anisotropy in $\omega_c\tau$ (the
product of the cyclotron frequency and the transport lifetime).

In order to extract information on the FS and $\omega_c\tau$, we carried out a least-square fitting of the data using
the Shockley-Chambers tube integral form of the Boltzmann transport equation modified for a quasi-2D metal with a
four-fold anisotropic scattering rate 1/$\tau(\varphi)$ = $(1+\alpha\cos{4\varphi})/\tau^0$ and anisotropic in-plane
velocity $v_F(\varphi)$, incorporated via $1/\omega_c(\varphi) =(1+\beta\cos{4\varphi})/\omega_c^0$ \cite{Majed06,
Kennett06}. The sign of $\alpha$ defines the location of maximal scattering. The FS wavevector $k_F(\theta,\varphi)$
was parameterized by the lowest-order harmonic components satisfying the body-centered-tetragonal symmetry of Tl2201
\cite{BergemannAP},
\begin{eqnarray}
&&k_F(\theta,\varphi)=k_{00} - k_{40}\cos{4\varphi} - k_{21}\cos{(k_z c/2)}\sin{2\varphi}\nonumber\\
&& - k_{61}\cos{(k_z c/2)}\sin{6\varphi} - k_{101}\cos{(k_z c/2)}\sin{10\varphi}
 \end{eqnarray}
where $k_z$ is the $c$-axis wavevector and $c$ the interlayer spacing. The eight-fold term $k_{80}$cos(8$\varphi$) had
a negligible effect on the parameterization and was therefore not included in the fitting procedure. Note that the
$c$-axis warping parameters $k_{21}$, $k_{61}$ and $k_{101}$ are small compared to $k_{00}$, the radius of the
cylindrical FS (about the zone corners), and $k_{40}$, its in-plane squareness, and only ratios (e.g. $k_{61}/k_{21}$)
can be determined to good accuracy. To minimize the number of free parameters, we fix $k_{101}/k_{21}$ =
$k_{61}/k_{21}$ - 1 such that $t_{\perp}$($\varphi$) vanishes at $\varphi = 0^{\rm o}$ and 45$^{\rm o}$ \cite{Hussey03,
Andersen95} and fix $k_{00}$ using the empirical relation $T_c$/$T_c^{\rm max} = 1 - 82.6(p-0.16)^2$ with $T_c^{\rm
max}$ = 92K  and $(\pi k_{00}^2)/(2\pi/a)^2 = (1+p)/2$ \cite{Presland91}. $\beta$ depends largely on our choice of
$k_{61}/k_{21}$ with the best least-square values giving $\beta = 0 \pm 0.1$ for 0.6 $\leq$ $k_{61}/k_{21}$ $\leq$ 0.8
for all samples \cite{Analytis07}. The sum $\alpha + \beta$ was much less sensitive to variations in $k_{61}/k_{21}$
and for simplicity, we assume hereafter $\omega_c$($\varphi$) = $\omega_c^0$. For completeness however, in Table
\ref{tab:table1} we list the values for $\alpha + \beta$.

\begin{table}[t]
\caption{\label{tab:table1} Parameters obtained from the least-square fitting of AMRO data for six different Tl2201
crystals at 40K. Errors given in parentheses determine the error bars in Fig. 2.} \vspace{0.3cm}
\begin{ruledtabular}
\begin{tabular}{ccccccc}

$T_c$ & $k_{00} (\AA^{-1})$ & $k_{40} (\AA^{-1})$ & $k_{61}/k_{21}$ & $\omega_c^0\tau^0$ & $\alpha$ + $\beta$ \\
\hline
15K & 0.730 & 0.038(3) & 0.73(8) & 0.27(1) & 0.31(2)\\
15K & 0.730 & 0.034(3) & 0.71(13) & 0.28(2) & 0.32(3)\\
17K & 0.729 & 0.035(3) & 0.68(12) & 0.29(2) & 0.29(3)\\
20K & 0.728 & 0.033(3) & 0.67(10) & 0.26(2) & 0.36(4)\\
32K & 0.726 & 0.037(3) & 0.67(10) & 0.21(2) & 0.42(3)\\
35K & 0.725 & 0.030(3) & 0.71(11) & 0.21(2) & 0.45(3)\\
\end{tabular}
\end{ruledtabular}

\end{table}

The best fits, shown in the panels to the right of each data set, are all excellent and the four remaining fitting
parameters displayed in Table \ref{tab:table1} appear well constrained due to the wide range of polar and azimuthal
angles studied \cite{Tl15Ka}. Within our experimental resolution, the FS parameters appear to have negligible doping
dependence. Moreover, the projected in-plane FS is found to be in good agreement with a recent ARPES study on the same
compound \cite{Plate05}. The anisotropy parameter $\alpha$ increases with rising $T_c$ whilst $\omega_c^0\tau^0$ shows
the opposite trend (reflecting the overall reduction in the ADMR in Fig. \ref{fig1}).

\begin{figure}[h]
\begin{center}
\includegraphics[width=7.8cm,keepaspectratio=true]{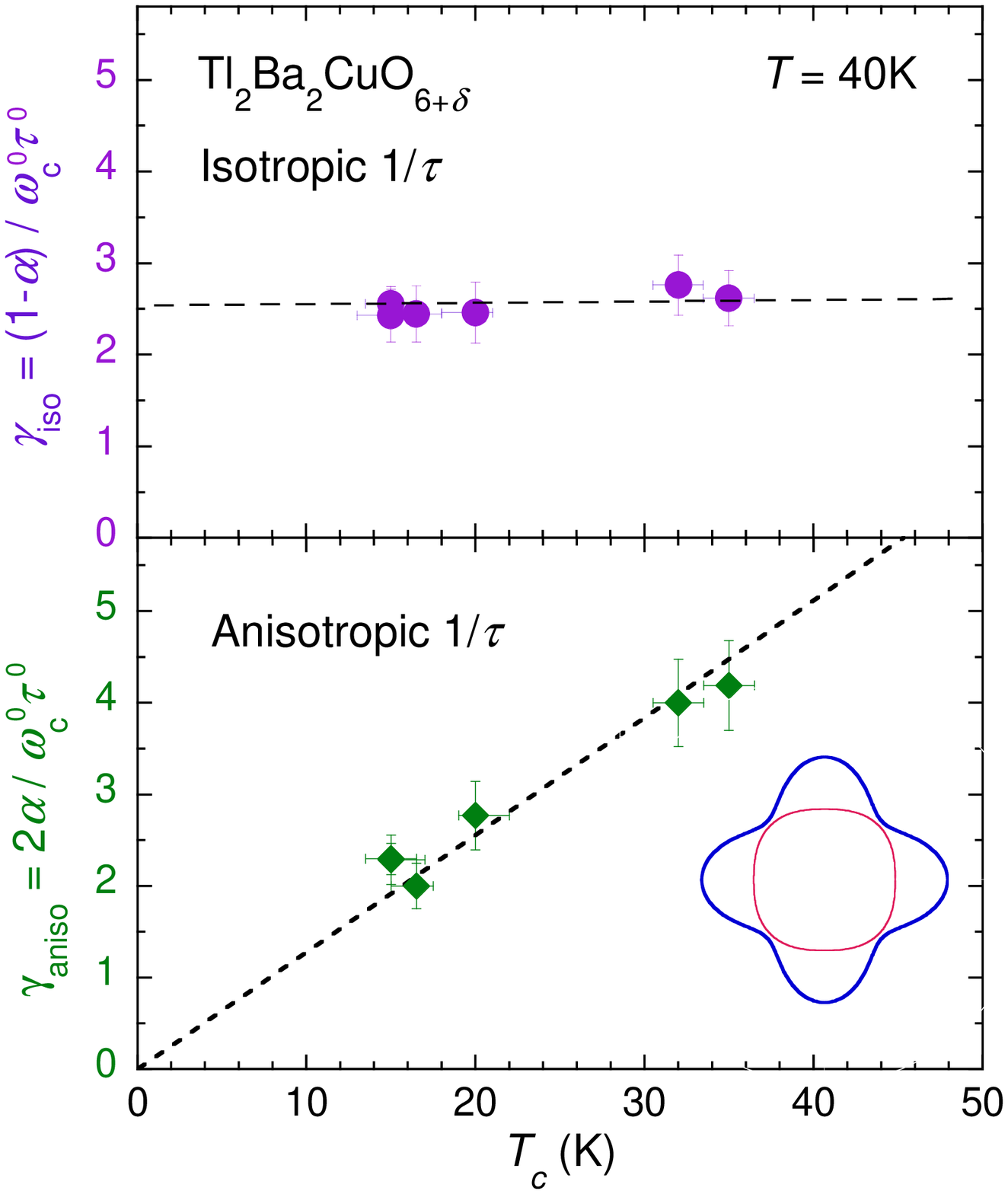}
\caption{(Color online) Top. Doping dependence of $\gamma_{\rm iso} \equiv (1-\alpha)/\omega_c^0\tau^0$ at $T$= 40K.
Bottom. Doping dependence of $\gamma_{\rm aniso} \equiv 2\alpha/\omega_c^0\tau^0$ at $T$ = 40K and $\varphi$ = 0. Here
we have assumed $\beta$ = 0. For $\beta$ = $\pm$ 0.1 (our range of uncertainty in $\beta$), the $\gamma_{\rm iso}$
($\gamma_{\rm aniso}$) points are shifted by $\pm 0.4$ ($\mp 0.4$) respectively. Inset: In-plane variation of
$\gamma_{\rm aniso}$ (thick blue line) with respect to $k_F(\varphi)$ (thin red line). \label{fig2}}
\end{center}
\end{figure}

Our previous analysis of Tl15Ka implied the existence of both isotropic and anisotropic components in the scattering
rate in OD Tl2201 \cite{Majed06}. Accordingly we split $1/\omega_c^0\tau$($\varphi$) into two components $\gamma_{\rm
iso}=(1-\alpha)/\omega_c^0\tau^0$ and $\gamma_{\rm aniso} = 2\alpha/\omega_c^0\tau^0$cos$^22\varphi$. In Tl15Ka,
$\gamma_{\rm iso} \sim A + BT^2$, due to a combination of impurity and electron-electron scattering, whilst
$\gamma_{\rm aniso} \sim CT$ (setting $\beta$ = 0) \cite{Majed06}, the microscopic origin of which has yet to be
identified. The doping ($T_c$) dependence of $\gamma_{\rm iso}$ and $\gamma_{\rm aniso}$ (at $T$ = 40K) are shown in
the top and bottom panels of Fig. \ref{fig2} respectively. Although we cannot say anything about the $T$-dependence of
the two components, it is clear that whilst $\gamma_{\rm iso}$ is essentially doping-independent, $\gamma_{\rm aniso}$
shows a clear linear scaling with $T_c$, extrapolating to zero at the onset of superconductivity (on the OD side).
Within a standard rigid band model, one would expect anisotropy in $v_F$$(\varphi)$ and hence in $\omega_c(\varphi)$ to
{\it increase} with doping as the FS at the Brillouin zone boundary approaches the saddle points. The fact that the
data show the opposite trend is therefore significant and justifies our key assumption that the basal-plane anisotropy
in Tl2201 is dominated by anisotropy in 1/$\tau(\varphi)$ \cite{Majed06, Analytis07}, which, surprisingly for such
highly OD samples, remains significant (the absolute anisotropy $\geq 2$ for Tl35K at $T$ = 40K).

This empirical correlation between $\gamma_{\rm aniso}$ and $T_c$ has several implications for the normal state of OD
cuprates. Firstly, given that the carrier concentration is varying only weakly (as $1+p$) in this doping range, it is
apparent that the marked change in the absolute value of resistivity with doping \cite{Kubo91} is due primarily to a
decrease in {\it anisotropic} scattering. Secondly, the observation of cyclotron motion over the full FS volume implies
that there are no pseudogapped regions of FS in OD Tl2201 for $x \geq 0.24$ {\cite{TallonLoram}. Whilst one might argue
that such a large applied field could suppress any pseudogapping in our crystals \cite{Shibauchi01}, our conclusion is
consistent with the observation of metallic $\rho_{\bot}$($T$) in zero-field \cite{Hussey03, Hussey96}. The $c$-axis
pseudogap, inferred from interlayer tunnelling \cite{Shibauchi01} and scanning tunnelling microscopy (STM)
\cite{Renner98} on Bi$_2$Sr$_2$CaCu$_2$O$_{8+\delta}$, is therefore a non-universal feature of OD cuprates.

Lastly and most importantly, the suppression of superconductivity on the OD side looks to coincide with the
disappearance of $\gamma_{\rm aniso}$. This conclusion is consistent with the observation in non-superconducting
La$_{1.7}$Sr$_{0.3}$CuO$_4$ of a strictly $T^2$ resistivity \cite{Nakamae03} (provided $\gamma_{\rm aniso}$ remains
$T$-linear) and a vanishing of $T$-dependence of the resistive anisotropy $\rho_{\bot}/\rho_{ab}$($T$)
\cite{Nakamura93, IoffeMillis} and indicates an intimate relation between the anisotropic scattering and HTSC. Our
findings contrast markedly with the doping evolution of the imaginary part of the quasiparticle self-energy Im$\Sigma$
inferred from ARPES. Here the nodal/antinodal quasiparticle anisotropy is seen to vanish \cite{Yang06} or even reverse
its sign \cite{Zhou04, Plate05} before superconductivity is suppressed on the OD side. ARPES-specific issues such as
resolution, background subtraction and matrix-element effects are not believed to be important here \cite{Peets06}. One
thus needs to ask the question whether the two probes are in fact measuring the same quantity. Direct comparison of the
scattering rates deduced from ADMR (this study and Ref. \cite{Hussey96}) and ARPES \cite{Plate05} in Tl2201 ($T_c \sim$
30K) shows that the former is more than one order of magnitude smaller. Since ARPES measures the quasiparticle
lifetime, small-angle scattering must contribute significantly to Im$\Sigma$. Interlayer transport in cuprates however
is also believed to be determined by a product of {\it single-particle} spectral functions on adjacent planes
\cite{Sandemann01} and may thus be similarly susceptible to small-angle scattering. Given that ADMR is a bulk probe, we
conjecture that the anomalously large linewidths seen in ARPES most probably arise from additional scattering at the
(cleaved) surface.

\begin{figure}[h]
\begin{center}
\includegraphics[width=7.8cm,keepaspectratio=true]{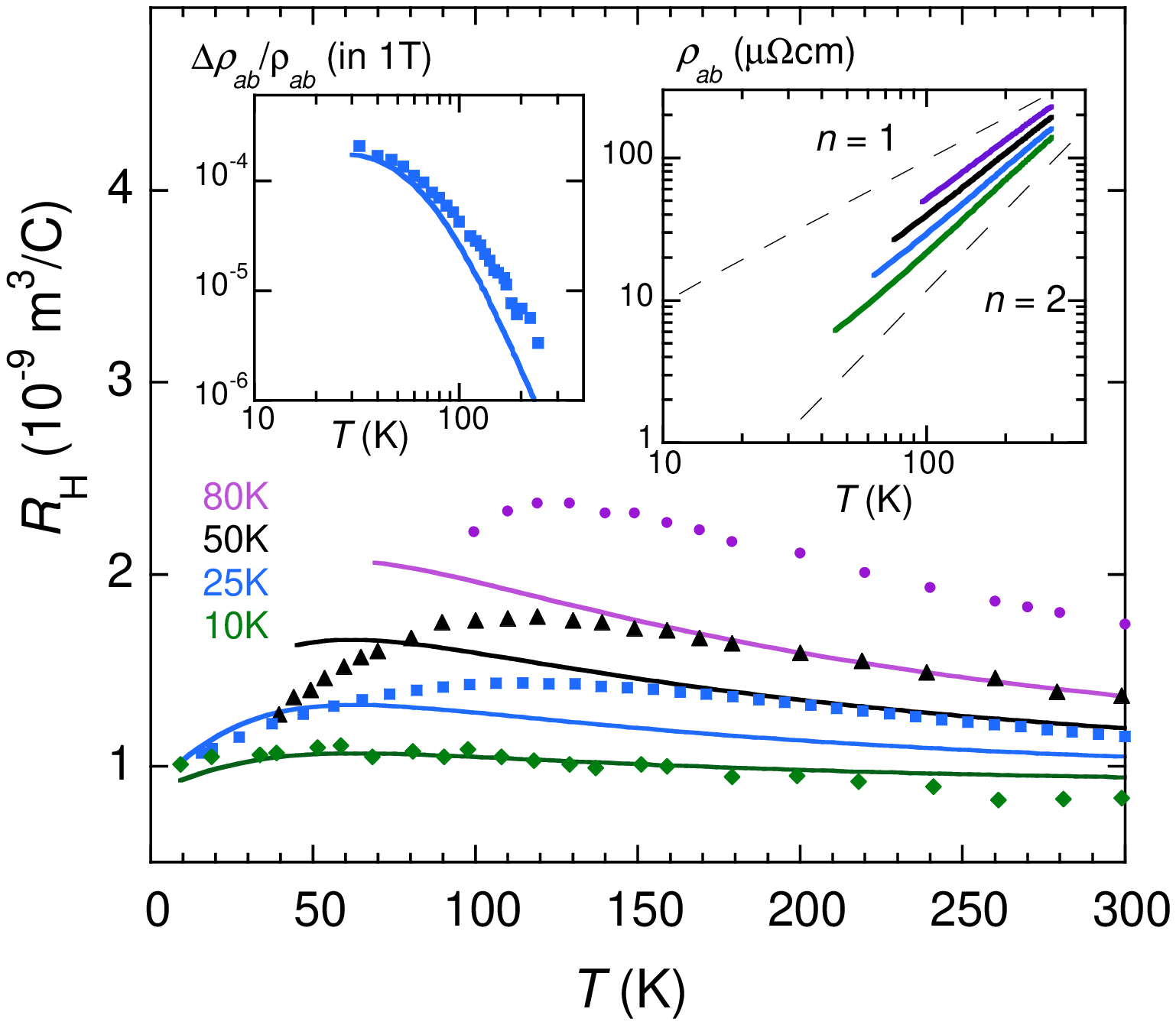}
\caption{(Color online) Main. $R_{\rm H}(T)$ simulations for Tl2201 with $T_c$ values of 10, 25, 50 and 80K (solid
lines) together with published data for $10$K (green diamonds \cite{Manako92}), 25K (blue squares \cite{Hussey96})and
50K (black triangles \cite{Manako92}) single crystals plus polycrystalline data for $T_c$ = 81K (purple circles,
\cite{Kubo91}). Top right. Simulation of $\rho_{ab}(T) - \rho_{ab}(T=0)$ (on a log-log scale) using the same
parameterization. Top left. Solid line: Simulation of $\Delta \rho_{ab}/\rho_{ab}(T)$ at 1T for $T_c$ = 25K. Blue
squares: $\Delta \rho_{ab}/\rho_{ab}(T)$ data at 1T for $T_c$ = 25K crystal \cite{Hussey96}. \label{fig3}}
\end{center}
\end{figure}

As stated above, $\gamma _{\rm aniso}(T)$ could account for both $R_{\rm H}$($T$) and $\rho_{ab}(T)$ in Tl15Ka at low
$T$ \cite{Majed06}. To simulate the {\it doping} evolution of $\rho_{ab}$($T,p$) and $R_{\rm H}$($T,p$) in Tl2201, we
adopt a simple one-parameter-scaling model and calculate $\rho_{ab}$, $R_{\rm H}$ (and the in-plane magnetoresistance
$\Delta\rho_{ab}/\rho_{ab}$) using the Jones-Zener form of the Boltzmann equation for a quasi-2D FS \cite{Hussey03b}.
Firstly, $k_{40}$ and $k_{00}$ are fixed using the values for Tl15Ka listed in Table \ref{tab:table1} and the scaling
relation \cite{Presland91} described earlier. Secondly, we assume that 1/$\omega_c\tau$($T, \varphi$) has the same form
as Tl15Ka for all crystals, i.e. 1/$\omega_c\tau$($T, \varphi$) = $\gamma_{\rm iso}$($T$) + $\gamma_{\rm
aniso}$($T,\varphi$) = $A + BT^2 + C(p)T$cos$^22\varphi$ where $C(p)$ = $C$(Tl15Ka) $\times$ $T_c(p)/15$
\cite{Majed06}. Finally, the anticipated return to isotropic scattering at high $T$ is simulated by inclusion (in
parallel) of a maximum scattering rate $\Gamma_{\rm max}$ = $\langle v_F(\varphi) \rangle/a$ (where $a$ is the in-plane
lattice spacing) in accord with the Ioffe-Regel limit \cite{Hussey03b}. In this way, $\gamma(T,\varphi)$ saturates at
different points on the FS at different $T$. Such anisotropy in the onset of saturation (of Im$\Sigma(\omega)$) has
been seen in optimally doped La$_{2-x}$Sr$_x$CuO$_4$ \cite{Chang06}.

The resultant simulations for $R_{\rm H}$($T,p$) and $\rho_{ab}$($T,p$) are presented in Fig. \ref{fig3} for $T_c$ =
10, 25, 50 and 80K. Despite there being no free parameters ($C(p)$ is fixed by $T_c$), the $R_{\rm H}$($T$) plots show
qualitative agreement with the published data (solid symbols, \cite{Kubo91, Hussey96, Manako92}). The $p$-dependence of
$\rho_{ab}(T)$ is also consistent with experiment, where the exponent $n$ of $\rho_{ab}$ (= $\rho_0 + \nu T^n$) is
found to evolve smoothly from 1 to 2 between optimal doping and the SC/non-SC boundary \cite{Kubo91}. Finally,
$\Delta\rho_{ab}/\rho_{ab}$ (top left panel for $T_c$ = 25K) displays the correct magnitude and $T$-dependence
\cite{Hussey96}. This correspondence, plus the fact that our simulations capture the marked increase in the magnitude
of $R_{\rm H}$ for what are relatively small changes in carrier concentration, suggests that anisotropic scattering is
a dominant contributor to the $T$- and $p$-dependence of $R_{\rm H}$ across the OD regime, whilst the {\it monotonic}
variation of $R_{\rm H}$($T$, $p$) is incompatible with a sign reversal of the scattering rate anisotropy near $p$ =
0.2 inferred from ARPES \cite{Zhou04, Plate05}.

As for possible origins of the observed anisotropy, real-space (correlated) electronic inhomogeneity \cite{McElroy05}
is one possible candidate though as yet, no measurements have been performed on heavily OD non-SC cuprates to establish
any possible link between inhomogeneity and superconductivity. Interactions with a bosonic mode are another
possibility, though given the strong angle and doping dependence, presumably not phonons. More likely candidates
include spin, charge (stripe) or $d$-wave pairing fluctuations which all disappear with superconductivity on the OD
side \cite{Wakimoto04, Reznik06, IoffeMillis}. The preservation of the $T$-linear scattering rate to low $T$ though
implies a vanishingly small energy scale for such fluctuations, characteristic of proximity to a quantum critical
point. Intriguingly, the transport scattering rate close to a 2D Pomeranchuk instability was recently shown to have a
form identical to that observed in OD Tl2201 \cite{Dell'Anna06}. In such a scenario however, $\gamma_{\rm aniso}$
should strengthen, rather than diminish, with doping as the van Hove singularity is approached on the OD side.

In conclusion, we have found a correlation between $T_c$ and the anisotropy of the transport scattering rate in OD
Tl2201. Although necessarily focused on heavily OD cuprates, this observation has implications for the entire phase
diagram, in particular the evolution of $R_{\rm H}$($T,p$). In addition, the present results support a previously
unforeseen link between the normal state scattering and superconducting mechanisms. Finally our ADMR results affirm
that maximal scattering remains at the antinodes throughout the OD region, in contrast to what has been recently
inferred from ARPES measurements.

\begin{acknowledgments}
We thank A. Damascelli, M. P. Kennett, R. H. McKenzie and J. A. Wilson for helpful discussions. This work was supported
by EPSRC and a co-operative agreement between the State of Florida and NSF.
\end{acknowledgments}

\bibliography{apsmajed}% Produces the bibliography via BibTeX.

\begin{thebibliography}{32}
\bibitem{asymmetry} The asymmetry around $\theta$ = 0 arises from slight misalignment of the crystalline $c$-axis with respect to the
rotational platform. The inclusion of a coordinate axis rotated around the platform axis can account for such a
distortion and has been added to our fitting routine though has a negligible effect on the parameters obtained.
\bibitem{Tl15Ka} Note for Tl15Ka, top line in Table \ref{tab:table1}, data has been obtained for only one azimuthal angle
\cite{Majed06}.
\end{thebibliography}

\end{document}